# An Algorithm to Reduce the Time Complexity of Earliest Deadline First Scheduling Algorithm in Real-Time System

Jagbeer Singh

Dept. of Computer Science and Engineering, Gandhi Institute of Engineering and Technology Gunupur, Rayagada (Orissa), India-765022,
E-mail:- willybokadia@gmail.com , Mob No. +919439286506, +917873887979

## ABSTRACT

In this paper I have study to Reduce the time Complexity of Earliest Deadline First (**EDF**), a global scheduling scheme for Earliest Deadline First in Real Time System tasks on a Multiprocessors system. Several admission control algorithms for Earliest Deadline First (**EDF**) are presented, both for hard and soft real-time tasks. The average performance of these admission control algorithms is compared with the performance of known partitioning schemes. I have applied some modification to the global Earliest Deadline First (**EDF**) algorithms to decrease the number of task migration and also to add predictability to its behavior. The Aim of this work is to provide a sensitivity analysis for task deadline context of multiprocessor system by using a new approach of **EFDF** (Earliest Feasible Deadline First) algorithm. In order to decrease the number of migrations we prevent a job from moving one processor to another processor if it is among the **m** higher priority jobs. Therefore, a job will continue its execution on the same processor if possible (processor affinity). The result of these comparisons outlines some situations where one scheme is preferable over the other. Partitioning schemes are better suited for hard real-time systems, while a global scheme is preferable for soft real-time systems.

## INTRODUCTION

In Earliest Deadline First (**EDF**) scheduling, at every scheduling point the task having the shortest deadline is taken up for scheduling. The basic principle of this algorithm is very intuitive and simple to understand. The schedulability test for **EDF** is also simple. A task is schedule under **EDF**, if and only if it satisfies the condition that total processor utilization (*u<sub>j</sub>*) due to the task set is less than **1**.

With scheduling periodic processes that have deadlines equal to their periods, **EDF** has a utilization bound of 100%. Thus, the schedulability test for **EDF** is:

$$U = \sum_{i=1}^{n} \frac{C_i}{T_i} \leq 1,$$

where the $\{C_i\}$ are the worst-case computation-times of the *n* processes and the $\{T_i\}$ are their respective inter-arrival periods (assumed to be equal to the relative deadlines).

For example let us Consider 3 periodic processes scheduled using **EDF**, the following acceptance test shows that all deadlines will be met.

| Process | Execution Time = C | Period = T |
|---------|--------------------|------------|
| P1      | 1                  | 8          |
| P2      | 2                  | 5          |
| P3      | 4                  | 10         |

The utilization will be:

$$\frac{1}{8} + \frac{2}{5} + \frac{4}{10} = 0.925 = 92.5\%$$

The theoretical limit for any number of processes is 100% and so the system is schedulable.

**EDF** has been proven to be an optimal uniprocessor scheduling algorithms [8].This means that if a set of tasks is unschedulable under **EDF**, then no other scheduling algorithm can feasible schedule this task set. The **EDF** algorithm chooses for execution at each instant in the time currently active job(s) that have the nearest deadlines. The **EDF** implementation upon uniform parallel machines is according to the following rules [2], No Processor is idled while there are active jobs waiting for execution, when fewer then *m* jobs are active, they are required to execute on the fastest processor while the slowest are idled, and higher priority jobs are executed on faster processors.

A formal verification which guarantees all deadlines in a real-time system would be the best. This verification is called feasibility test.

Three different kinds of tests are available:-

- Exact tests with long execution times or simple models [11], [12], [13].
- Fast sufficient tests which fail to accept feasible task sets, especially those with high utilizations [14], [15].
- Approximations, which are allowing an adjustment of performance and acceptance rate [1], [8].

For many applications an exact test or an approximation with a high acceptance rate

Must be used. For many task sets a fast sufficient test is adequate.

**EDF** is an appropriate algorithm to use for online scheduling on uniform multiprocessors. However, their implementation suffers from a great number of migrations due to vast fluctuations caused by finishing or arrival of jobs with relatively nearer deadlines. Task migration cost might be very high. For example, in loosely coupled system such as cluster of workstation a migration is performed so slowly that the overload resulting from excessive migration may prove unacceptable [3]. Another disadvantage of **EDF** is that its behavior becomes unpredictable in overloaded situations. Therefore, the performance of **EDF** drops in overloaded condition such that it cannot be considered for use. In this paper I am presenting a new approach, call the Earliest Feasible Deadline First (**EFDF**) which is used to reduce the time complexity of earliest deadline first algorithm by some assumptions.

## REVIEW OF RELATED WORK

Each processor in a uniform multiprocessor machine is characterized by a speed or Computing capacity, with the interpretation that a job executing on a processor with speed $s$ for $t$ time units completes ($s * t$) units of execution. The earliest-deadline first (**EDF**) scheduling of real-time systems upon uniform multiprocessor machines is considered. It is known that online algorithms tend to perform very poorly in scheduling

Such real-time systems on multiprocessors; resource-augmentation techniques are presented here that permit online algorithms in general (**EDF** in particular) to perform better than may be expected given these inherent limitations. It is shown that **EDF** scheduling upon uniform multiprocessors is robust with respect to both job execution requirements and processor computing capacity.

## PROPOSED APPROACH

I have applied some modification to the global Earliest Deadline First (**EDF**) algorithms to decrease the number of task migration and also to add predictability to its behavior. In order to decrease the number of migrations we prevent a job from moving to another processor if it is among the **m** higher priority jobs. Therefore, a job will continue its execution on the same processor if possible (***processor affinity***[1]).

In Earliest Deadline First (**EDF**) scheduling, at every scheduling point the task having the shortest deadline is taken up for scheduling. The basic principle of this algorithm is very intuitive and simple to understand. The schedulability test for Earliest Deadline First (**EDF**) is also simple. A task is schedule under **EDF**, if and only if it satisfies the condition that total processor utilization due to the task set is less than 1. For a set of periodic real-time task $\{T_1, T_2, T_n\}$, **EDF** schedulibility criterion can be expressed as:-

$$\sum_{i=1}^{n} \frac{e_i}{p_i} = \sum_{i=1}^{n} u_i \leq 1$$

Where $e_i$ is the execution time, $p_i$ is the priority of task and $u_i$ is the average utilization due to the task $T_i$ and $n$ is the total number of task in set. **EDF** has been proven to be an optimal uniprocessor scheduling algorithm [8]. This means that if a set of task is unschedulable under Earliest Deadline First (**EDF**), then no other scheduling algorithm can feasible schedule this task set. In the simple schedulability test for **EDF** we assumed that the period of each task is the same as its deadline. However in practical problem the period of a task may at times be different from its deadline. In such cases, the schedulability test needs to be changed. If $p_i > d_i$, then each task needs $e_i$ amount of computing time every $\min(p_i, d_i)$ duration time. Therefore we can write:

$$\sum_{i=1}^{n} \frac{e_i}{\min(p_i, d_i)} \leq 1$$

However, if $p_i < d_i$, it is possible that a set of tasks is **EDF** schedulable, even when the task set fail to meet according to expression.

My motivation for exploiting **processor affinity** drive from the observation that, for much parallel application, time spent bringing data into the local memory or cache is significant source of overhead, ranging between *30%* to *60%* of the total execution time [3]. While migration is unavoidable in the global schemes, it is possible to minimize migration caused by a poor assignment of task to processors. By scheduling task on the processor whose local memory or cache already contains the necessary data, we can significantly reduce the execution time and thus overhead the system. It is worth mentioning that still a job might migrate to another processor when there are two or more jobs that were last executed on the same processor. A migration might also happen when the numbers of ready jobs become less than the number processors. This fact means that our proposed algorithm is a work conserving one. In order to give the scheduler a more predictable behavior we first perform a **feasibility check** to see whether a job has a chance to meet its deadline by using some exiting algorithm like Yao's [16]. If so, the job is allowed to get executed. Having known the deadline of a task and its remaining execution time it is possible to verify whether it has the opportunity to meet its dead line. More precisely, this verification can be done by examining a *task's laxity*[3]. The *laxity* of a real-time task $T_i$ at time $t$, $L_i(t)$, is defined as follows:-

$$L_i(t) = D_i(t) - E_i(t)$$

Where $D_i(t)$ is the dead line by which the task $T_i$ must be completed and $E_i(t)$ is the amount of computation remaining to be performed. In other words, Laxity is a measure of the available flexibility for scheduling a task. A laxity of $L_i(t)$ means that if a task $T_i$ is delayed at most by $L_i(t)$ time units, it will still has the opportunity to meet its deadline. A task with zero laxity must be scheduled right away and executed without preemption or it will fail to meet its deadline. A negative laxity indicates that the task will miss the deadline, no matter when it is possible picked up for execution. We call this novel approach the **Earliest Feasible Deadline First** (**EFDF**) [4] scheduling algorithm whose details are presented in the following section given below.

## EFDF SCHEDULING ALGORITHM

Let $m$ denote the number of processing nodes and $n$, ($n \geq m$) denote the number of
Available tasks in a uniform parallel real-time system. Let $s_1, s_2, \ldots s_m$ denote the computing capacity of available processing nodes indexed in a non-increasing manner: $s_j \geq s_{j+1}$ for all $j$, $1 < j < m$. We assume that all speeds are positive i.e. $s_j > 0$ for all j. In this section I am presenting five steps of *EFDF* algorithm. Obviously, each task which is picked for up execution is not considered for execution by other processors.

**1.** Perform a feasibility check to specify the task which has a chance to meet their deadline and put them in a set **A**, Put the remaining tasks in set **B**.

**2.** Sort both task sets **A** and **B** according to their deadline in a non-descending order. Let $k$ denote the number of tasks in set **A**, i.e. the number of tasks that have the opportunity to meet their deadline.

**3.** For all processor $j$, ($j \leq min(k,m)$) check whether a task which was last running on the $j_{th}$ processor is among the first $min(k,m)$ tasks of set **A**. If so assign it to the $j_{th}$ processor. At this point there might be some processors to which no task has been assigned yet.

**4.** For all $j$, ($j \leq min(k,m)$) if no task is assigned to the $j_{th}$ processor, select the task with earliest deadline from remaining tasks of set A and assign it to the $j_{th}$ processor. If $k \geq m$, each processor have a task to process and the algorithm is finished.

5. If $k < m$, for all $j$, ($k < j \leq m$) assign the task with smallest deadline from **B** to the $j_{th}$ processor. The last step is optional and all the tasks from **B** will miss their deadlines.

## TIME COMPLEXITY

The **EDF** algorithm would be maintain all tasks that are ready for execution in a queue. Any freshly arriving task would be inserted at the end of queue. Each task insertion will be achieved in *O(1)* or constant time, but task selection (to run next) and its deletion would require *O(n)* time, where n is the number of tasks in the queue. **EDF** simply maintaining all ready tasks in a sorted priority queue that will be used a heap data structure. When a task arrives, a record for it can be inserted into the heap in $O(log_2 n)$ time where $n$ is the total number of tasks in the priority queue. Therefore, the time complexity of Earliest Deadline First (**EDF**) is equal to that of a typical sorting algorithm which is $O(nlog_2 n)$. While in the **EFDF** the number of distinct deadlines that tasks is an application can have are restricted. In my approach, whenever a task arrives, its absolute deadline is computed from its release time and its relative deadline. A separate first in first out (FIFO) queue is maintained for each distinct relative deadline that task can have. The schedulers insert a newly arrived task at the end of the corresponding relative deadline queue. So tasks in each queue are ordered according to their absolute deadlines. To find a task with the earliest absolute deadline, The scheduler needs to search among the threads of all FIFO queues. If the number of priority queue maintained by the scheduler in $n$, then the order of searching would be $O(1)$. The time to insert a task would also be $O(1)$. So finally the time complexity of five steps of Earliest Feasible Deadline First (**EFDF**) are $O(n)$, $O(nlog_2 n)$, $O(m)$, $O(m)$, $O(m)$, respectively.

## CONCLUSION AND FUTURE WORK

This work focus on some modification to the global Earliest Deadline First (**EDF**) algorithms to decrease the number of task migration and also to add predictability to its behavior. Mainly Earliest Feasible Deadline First (**EFDF**) algorithms are presented the least complexity according to their performance analyzed. Experimental result of Earliest Feasible Deadline First (**EFDF**) algorithm reduced the time complexity in compression of Earliest Deadline First (**EDF**) algorithm on real time system scheduling for multiprocessor system and perform the feasibility checks to specify the task which has a chance to meet their deadline. When Earliest Feasible Deadline First (**EDF**) is used to schedule a set of real-time tasks, unacceptable high overheads might have to be incurred to support resource sharing among the tasks without making tasks to miss their respective deadlines, due to this it will take again more time. My future research will investigate other less complexity Algorithm and also reduced the overhead for different priority assignments for global scheduling which will, consequently, lead to different bounds.

---

1. **Processor affinity** is a modification of the native central queue scheduling algorithm in a symmetric multiprocessing operating system. Each task (be it process or thread) in the queue has a tag indicating its preferred / kin processor. At allocation time, each task is allocated to its kin processor in preference to others.

## REFERENCES:-

**1.** S. Baruah, S. Funk, and J. Goossens , "*Robustness Results Concerning EDF Scheduling upon Uniform Multiprocessors*",*IEEE Transcation on computers*, Vol. 52, No.9 pp. 1185-1195 September 2003.

**2.** E.P.Markatos, and T.J. LeBlanc, " *Load Balancing versus Locality Management in Shared-Memory Multiprocessors*", *The 1992 International Conference on Parallel Processing*, August 1992.

**3.** S. Lauzac, R. Melhem, and D. Mosses,"*Compression of Global and Partitioning Scheme for Scheduling Rate Monotonic Task on a Multiprocessor*", *The 10th EUROMICRO Workshop on Real-Time Systems, Berlin*,pp.188-195, June 17-18, 1998.

**4.** Vahid Salmani ,Mohsen Kahani , " *Deadline Scheduling with Processor Affinity and Feasibility Check on Uniform Parpllel Machines*", Seventh International Conference on Computer and Information Technology, *CIT.121.IEEE*,2007.

**5.** S. K. Dhall and C. L. Liu, "On a real-time scheduling problem. Operations Research", 26(1):127–140, 1978.

**6.** Y. Oh and S. Son. "*Allocating fixed-priority periodic tasks on multiprocessor systems*",
Real-Time Systems Journal, 9:207–239, 1995.

**7.** J. Lehoczky, L. Sha, and Y. Ding. "*The rate monotonic Scheduling: Exact characterization and average case behavior*", *IEEE Real-time Systems Symposium*, pages 166–171, 1989.

**8.** C.M. Krishna and Shin K.G. Real-Time Systems. Tata McGrawiHill,1997.

**9.** S. Chakraborty, S. Künzli, L. Thiele. *Approximate Schedulability Analysis*. 23rd IEEE Real-Time Systems Symposium (RTSS), IEEE Press, 159-168, 2002.

**10.** J.A. Stankovic, M. Spuri, K. Ramamritham, G.C. Buttazzo. *Deadline Scheduling for Real-Time Systems **EDF** and Related Algorithms*. Kluwer Academic Publishers, 1998.

**11.** S. Baruah, D. Chen, S. Gorinsky, A. Mok. *Generalized Multiframe Tasks.* The International Journal of Time-Critical Computing Systems, 17, 5-22, 1999.

**12.** S. Baruah, A. Mok, L. Rosier. *Preemptive Scheduling Hard-Real-Time Sporadic Tasks on One Processor*. Proceedings of the Real- Time Systems Symposium, 182-190, 1990.

**13.** K. Gresser. *Echtzeitnachweis Ereignisgesteuerter Realzeitsysteme*. Dissertation (in german), VDI Verlag, Düsseldorf, 10(286), 1993.

**14.** M. Devi. *An Improved Schedulability Test for Uniprocessor Periodic Task Systems*. Proceedings of the 15th Euromicro Conference on Real-Time Systems, 2003.

**15.** C. Liu, J. Layland. *Scheduling Algorithms for Multiprogramming in Hard Real-Time Environments*. Journal of the ACM, 20(1), 46-61, 1973